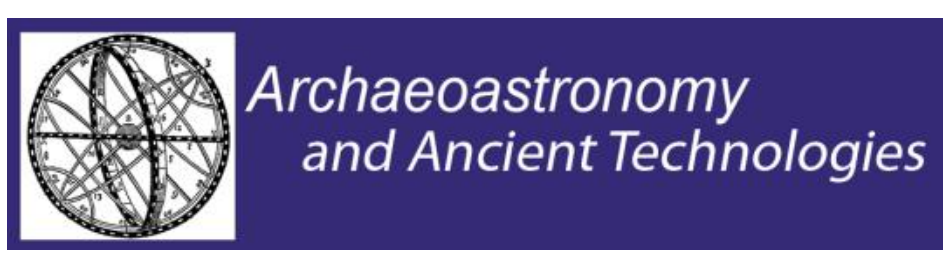

# Russian Meteorite of the Bronze Age (Rock Record)

**Larisa Vodolazhskaya [1],*, Mikhail Nevsky [2]**

[1] Department of Space Physics, Southern Federal University (SFU), str. Zorge, 5, Rostov-on-Don, 344090, Russian Federation; E-Mail: larisavodol@yahoo.com

[2] Astronomical Observatory SFU, Department of Space Physics, Southern Federal University (SFU), str. Zorge, 5, Rostov-on-Don, 344090, Russian Federation; E-Mail: munevsky@sfedu.ru

**Abstract**

This paper presents the results of the study of petroglyphs found in the quartzite grotto near the Skelnovsky small village in the Northern Black Sea in the South of Russia. The aim of the study was the analysis and interpretation of the Early Bronze Age petroglyphs using archaeoastronomical methods. The article presents a comparative analysis of Skelnovsky grotto ancient images and contemporary eyewitness accounts of the Sikhote-Alin meteorite fall and meteorite shower. Some petroglyphs were interpreted by us using ethnographic and folklore material. In this study, the magnetic declination for the geographical coordinates Skelnovsky farm was calculated, and the projection of the whole picture Skelnovskih petroglyphs on the topographical map of the area was built. The proposed location of the meteorite fall was determined with this projection. It is confirmed by satellite pictures, on which are the distinguishable terrain features, typical for the meteorite fall, are visible including the possible impact crater, and the corresponding symbols on a topographical map. The studies in the article conclude the astronomical character of the main content of the Skelnovsky petroglyphs picture which depicts the fall of a large meteoroid (bolide), similar to the Sikhote-Alin meteorite, accompanied by a meteorite shower.

After a comparative analysis of the images in Sklenovsky grotto it was discovered that the petroglyph is a copy of the figure depicted on the famous Mesopotamic clay tablet YBC 7289. We speculate that this figure is a prototype of an important building used either for ritual or farm purposes, typical for the Bronze Age and distributed as in Mesopotamia, as also on the coast of the Northern Black Sea. This may indicate a cultural continuity, including the field of protoscientific knowledge, between the inhabitants of the Northern Black Sea and the inhabitants of Mesopotamia during the Bronze Age.

## Introduction

Petroglyphs are found world-wide. Some of the images researchers interpret as signs of the Sun [1-5] and the Moon [6, 7]. Some of the petroglyphs, scientists are trying to associate with astronomical objects and phenomena: to associate with the starry sky [8, 9], with supernovae [10], with the appearance of comets [11-14], bolides [15, 16], meteors [17, 18], meteorite showers [19] and impact craters [20]. In 2009 the petroglyphs were discovered in the steppe zone of the south of the European part of Russia the first [21]. Basically, the petroglyphs were described by archaeologists as linear-geometric and did not get more interpretation. As part of our work, based on archaeoastronomical research methods, we propose to interpret the detected composition

Skelnovsky petroglyphs as detailed and fairly realistic portrayal of the large meteoroid fall, similar to the Sikhote-Alin meteorite, accompanied by a meteorite shower.

## Object of Study

The grotto with petroglyphs located on the right bank of the Don, 25 km from the Don, 6 km from the river Tikhaya near the Skelnovsky small village in the Verkhnedonskoy district of the Rostov region of Russia. Low slit-like grotto is in a quartzite block. Quartzite is characteristic of much of the Don and the Seversky Donets. Grotto has a rectangular shape and lies almost horizontally. Its base is a monolithic slab. Petroglyphs are depicted on it.

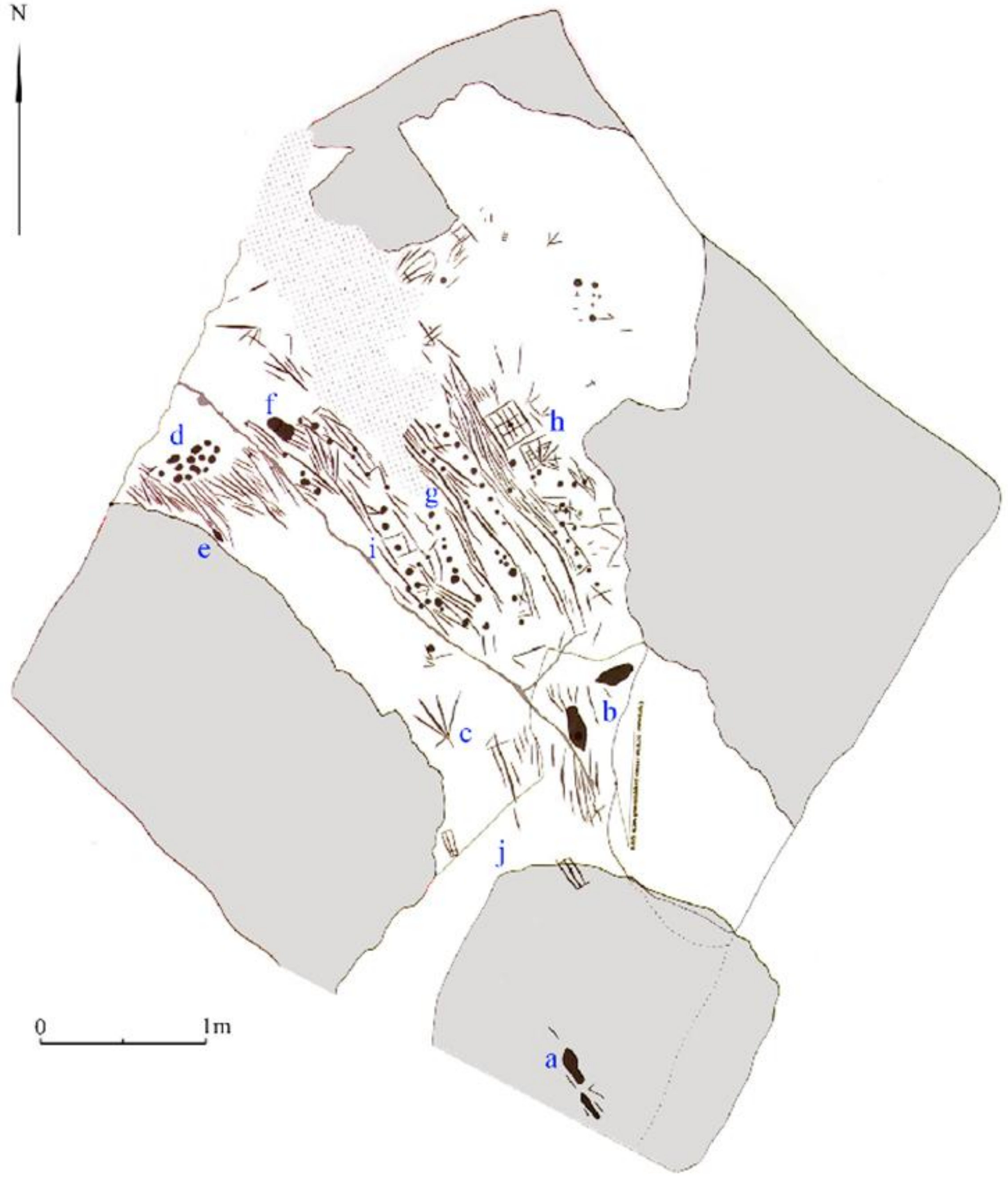

**Figure 1.** The plan is showing the location of petroglyphs in the grotto [21, fig. 12]. Petroglyphs: **a** - trail of human foot, **b** - stone axes, **c** - branch, **d** - meteorite field, **e** - spear, **f** - meteoroid, **g** - meteorite shower, **h** - large squares, **i** - small squares, **j** – rectangles.

Part of the slab at the north entrance has a strong natural damage, probably so it is not the petroglyphs were applied. The height of the cavity in the grotto is from 0.25 m to 0.8 m. In ancient times, all the petroglyphs were under the quartzite arch, some of which are now collapsed and the southern most of petroglyphs are closed fallen fragment of quartzite.

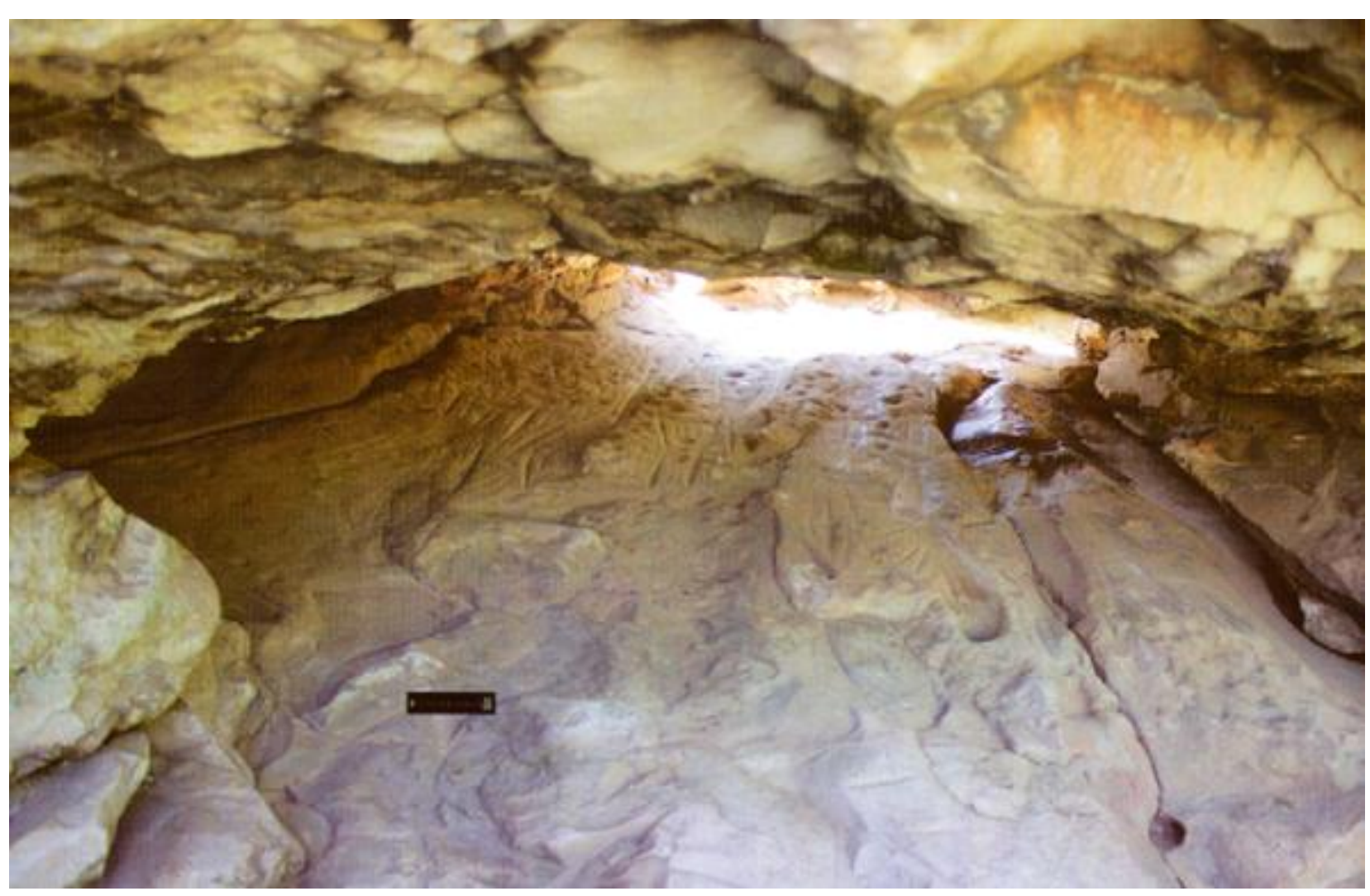

**Figure 2.** A general view of the grotto from the north [21, fig. 14].

Petroglyphs are applied in two ways: engraving and stationing. In the course of excavations at the northern entrance to the grotto were found engraved guns in the form of massive quartzite flakes and were found fragments of clay round bottom vessels that belong to Yamnaya culture (Early Bronze Age). V. J. Kiyashko dates Skelnovsky petroglyphs end of IV millennium BC [21]. Figure 1 schematically shows the location of the petroglyphs in the grotto. The direction of magnetic north is approximate. When measuring the direction of the north compass needle is constantly fluctuating[1]. A general view of the grotto is shown in Figure 2.

## Analysis of petroglyphs

We think that whole composition of petroglyphs is a holistic picture of the same event, and propose to consider the composition relative to the south-west entrance. There was found a petroglyph as a trace of the human foot, toward to the grotto entrance (Fig. 3).

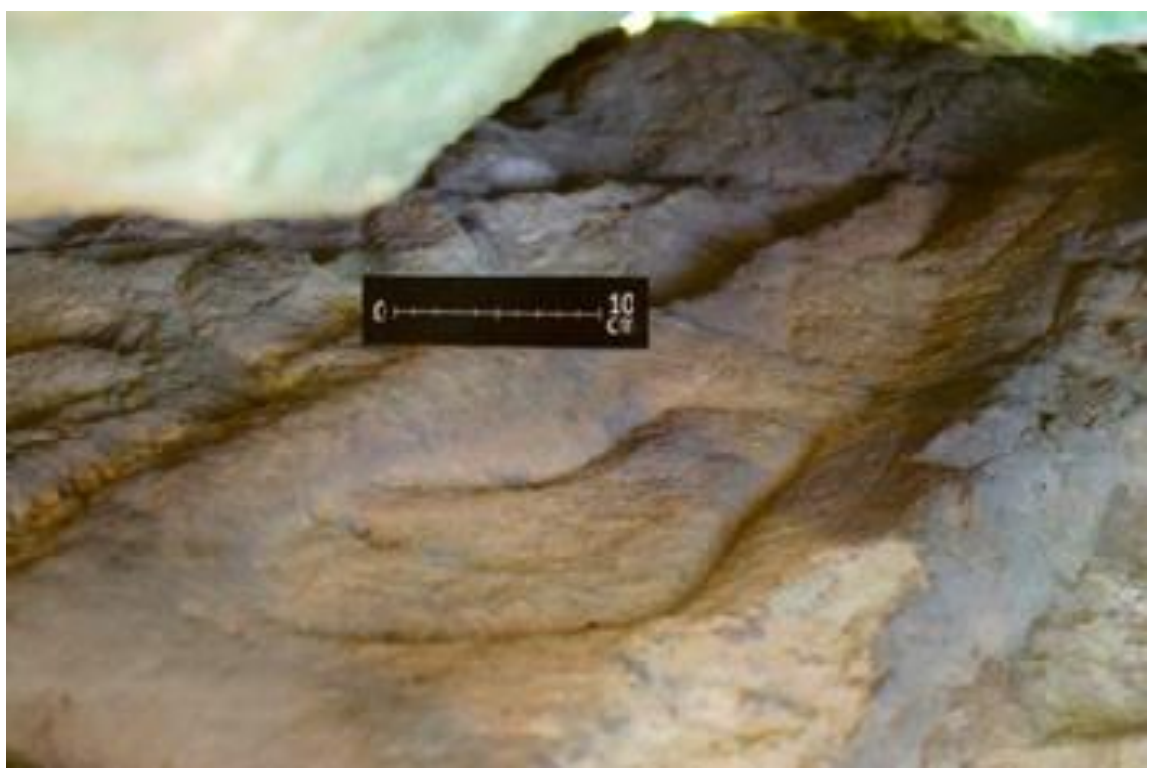

**Figure 3.** Petroglyph trail of the human foot [21, fig. 71].

[1] Information of V.Y. Kiyashko

At the entrance, to the north of the foot, realistically traced the outlines of two stone axes - hammers (Fig. 4a). Stationing pit at the central axes indicated drilled hole in the center (Fig. 4b).

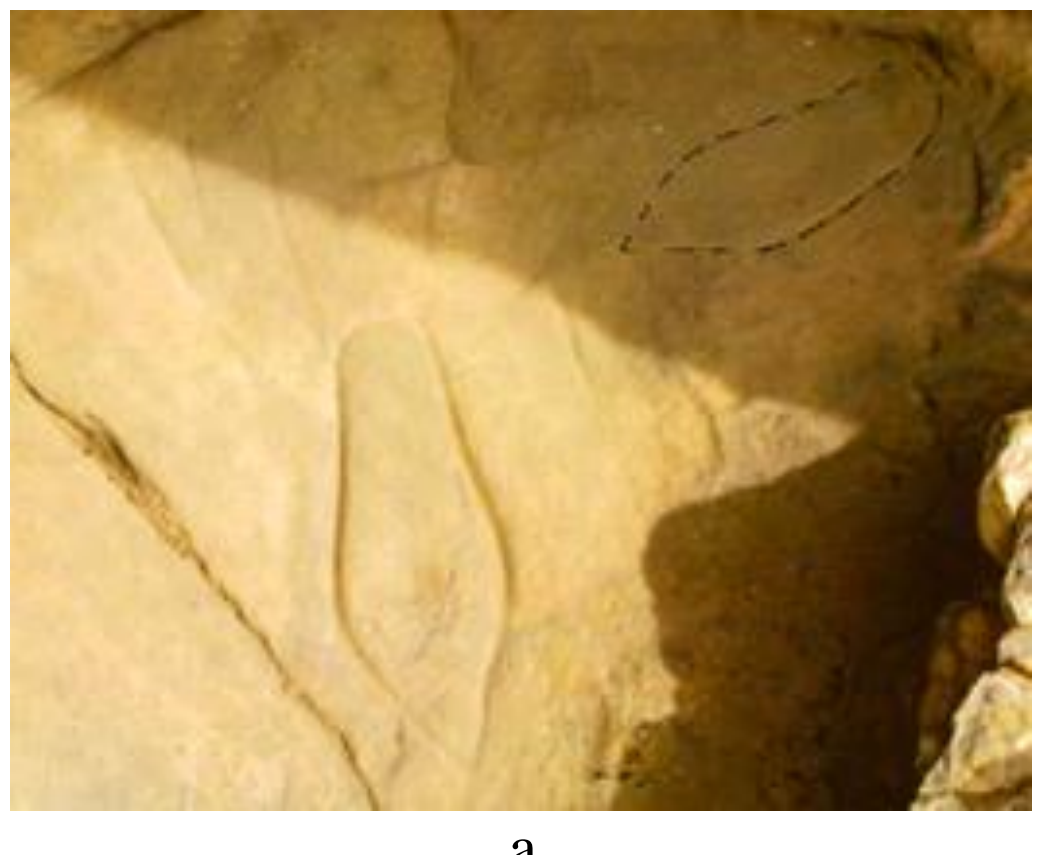

a

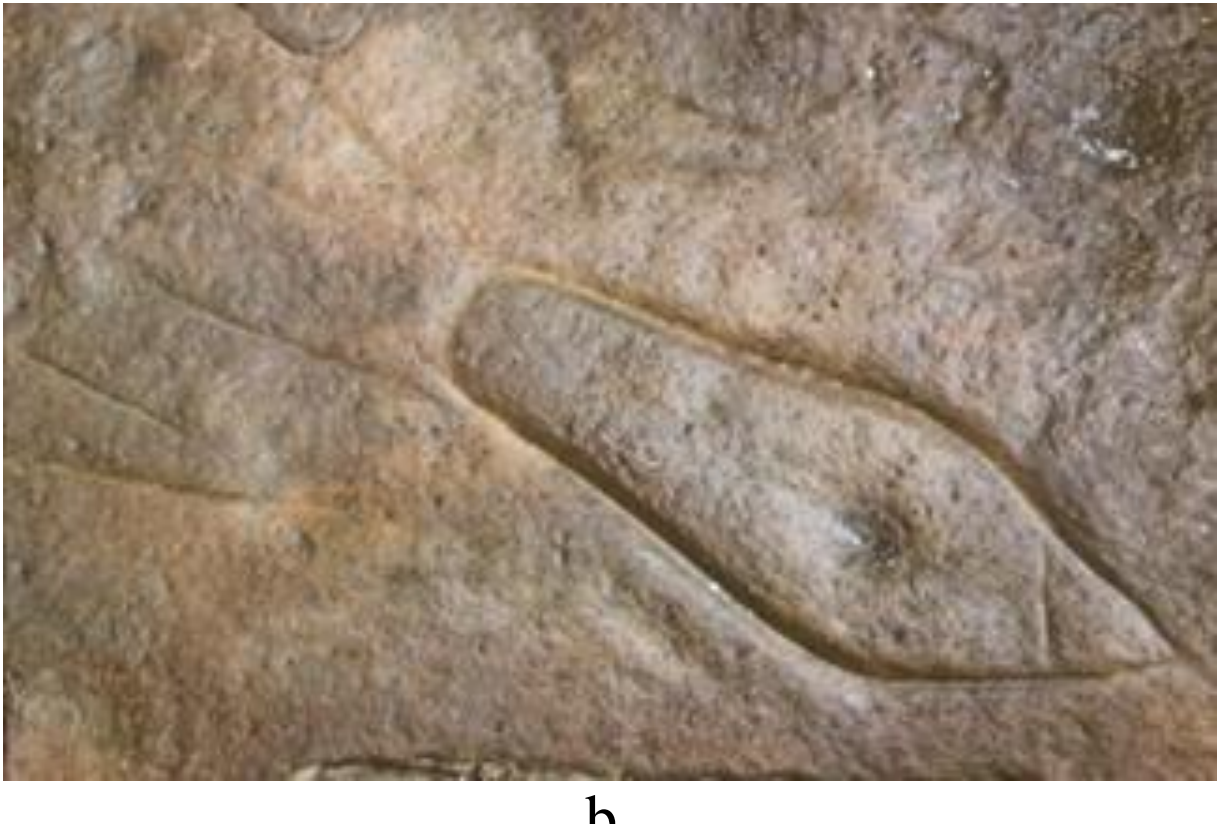

b

**Figure 4.** Petroglyphs of stone axes [21, fig. 66]. Petroglyph, which is in the shade, for clarity, circled in by the dotted line.

In Eastern Europe, drilled stone axes began to appear in the Eneolithic. In Tripolye archaeological sites found drilled stone axes with elongated butt similar in shape to the image of Skelnovsky grotto [22]. The sharp edge of a stone ax is directed nearly to the south - towards the trail foot. Emphasis is placed on the lines behind the stone axe. In conjunction with the orientation the axe in space North - South (the projection of the world axis), with its location in the center of the entrance, toward the trail foot (observer), the lines extending from the axe, can be interpreted as a symbol of the movement vertically falling stone axe. Eneolithic stone axes or hammers, according to the folklore of the Indo-European peoples are thunderstones and can symbolize meteorites [23]. Second, bad visibility petroglyph stone axe, located at right angles to the central axe, may indicate the direction of the appearance of thunderstone. We view the composition of petroglyphs in the grotto as a sketch of the events really took place and we believe that the second stone ax indicates that the meteoroid came from the east. The sky area near to zenith corresponds to the field of composition, located directly at the southern grotto entrance, and includes images of axes. In our view, this area of petroglyphs composition is a symbol of the complex - a kind of headline story recorded with petroglyphs in the depths of the grotto.

The petroglyph as branch located on a quartz plate in the zone associated with zenith too (Fig. 5).

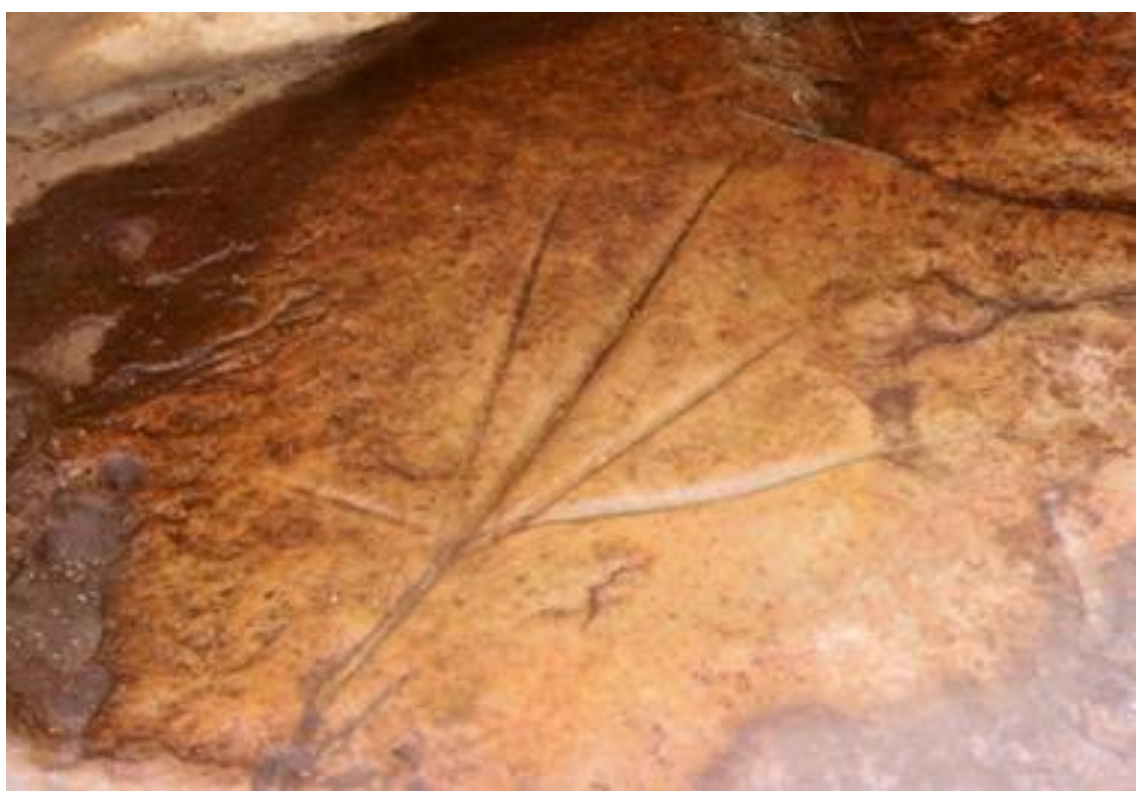

**Figure 5.** Petroglyph branch [21, fig. 56].

We consider that this petroglyph can illustrate visible occurrence of meteoroid in the sky and its first crushing with fragments flying apart like a fan. The lines, engraved on a slab of quartzite, depict luminous traces of flying fragments. In the fall of the Sikhote-Alin meteorite eyewitnesses said that the fireball appeared in the sky as bright stars, which flew a short distance, a dazzling flared, illuminating the surrounding countryside. Fireball is crushed several times after the first bright flash. At each crushing observed flashes of light [24, pp. 59-60].

Area, densely filled with cup-shaped pits, located in the northern part of the grotto in front of divergent lines of branch (Fig. 6). We interpret it as an image of meteorite field or part of it, and the pits, as the images of impact craters. From the pits crater field depart lines that may be indicated by falling meteorite fragments.

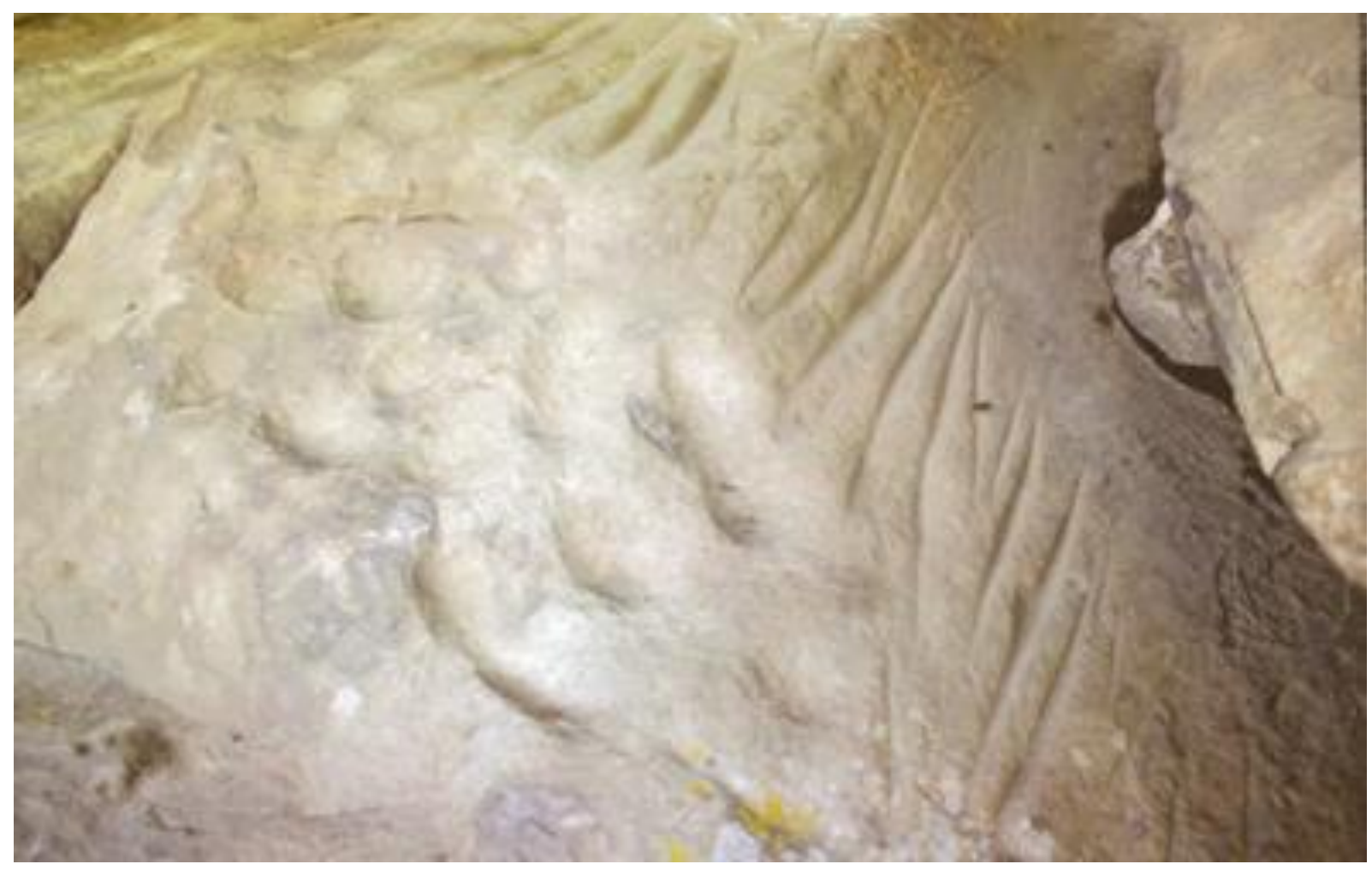

**Figure 6.** Cup-shaped pits in the northern part of the grotto [21, fig. 31].

In the area of Sikhote-Alin meteorite fall, according to witnesses, craters and funnels gaped among the debris of trees on the area of about one square kilometer. The largest of crater had a diameter of about 26 meters and a depth of 6 m. Trees, felled by the roots, lie radially around the craters within a radius of 20-30 m from the edges of the craters [24, p. 28]. The fall of the Sikhote-Alin meteorite occurred in the winter, and funnels stood out sharply on the white snow of their yellow-brown color, created with clay and stones. White quartzite of the Skelnovsky grotto evidence about the meteoroid fall in the winter time is also, possible. This could allow local people can easily see a large number of craters, and then capture them in a bowl-shaped petroglyphs in the grotto. Snow could allow local people to easily see a large number of craters, and then capture them in grotto cup-shaped petroglyphs.

One of the lines coming from the petroglyphs of meteor field pits ends with rhomboid petroglyphs that archaeologists have interpreted as the tip of the spear (Fig. 7). Depth of many lines in the composition is up to 2.0 cm and the depth of petroglyph tip lines is much less. We believe that it could be applied on a quartz plate well after the bulk of the petroglyphs, possibly in epoch of antiquity, as bolide in the Greek language means spear. For the ancient Greeks, there have been quite a realistic understanding of the meteoroid nature and characteristics of their fall. For example, Diogenes of Apollonia (V century BC) believed that meteorites are stones that rotate in space with the stars, they are hot, but invisible. He claimed that they often fall to the ground, leaving a fire trail and extinguish after the fall [25]. Anaxagoras (V century BC) are considered meteorites as fragments of red-hot stone mass of the Sun [26]. In addition, in the classical world was well-known fact of the fall of a large meteorite in Thrace near the Hellespont in 467 BC, which Anaxagoras predicted.

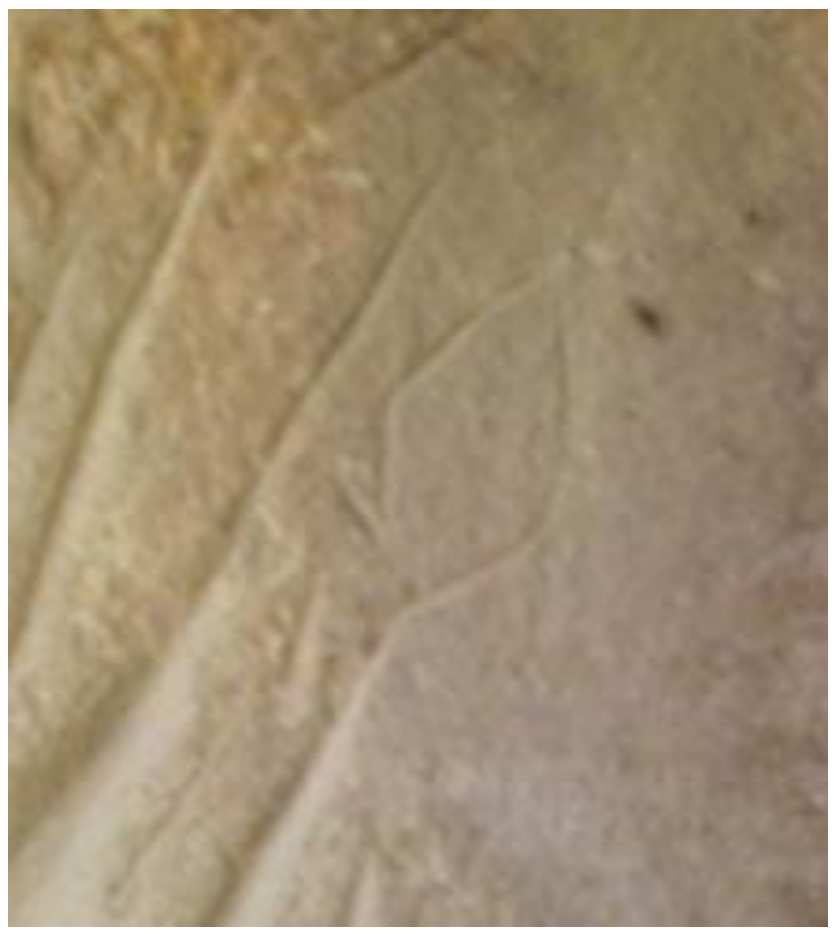

**Figure 7.** Petroglyph spear [21, fig. 58].

Before falling the meteoroid, from our point of view, has been portrayed by the ancient artist in the northern part of the grotto, next to petroglyphs of the meteorite field, but closer to the center of the northern entrance (Fig. 8). In the foreground of Figure 8 shows a pear-shaped pit, the biggest relative to other pits. It is the alleged image of a meteoroid. From the picture of the meteoroid to the south, deep into the grotto, moving away a few short lines similar to the lines coming from the image stone ax. Eyewitnesses of Sikhote-Alin meteorite fall, describing the meteoroid at the bottom of its trajectory, said that he was not spherical, but elongated - pear-shaped. They likened it to palm, mitten etc. [24, p. 59]. The cup-shaped pit in the Skelnovsky grotto, which we consider the image of a meteoroid, it has been interpreted by archaeologists as stylized palm or clawed paw [21, p. 8].

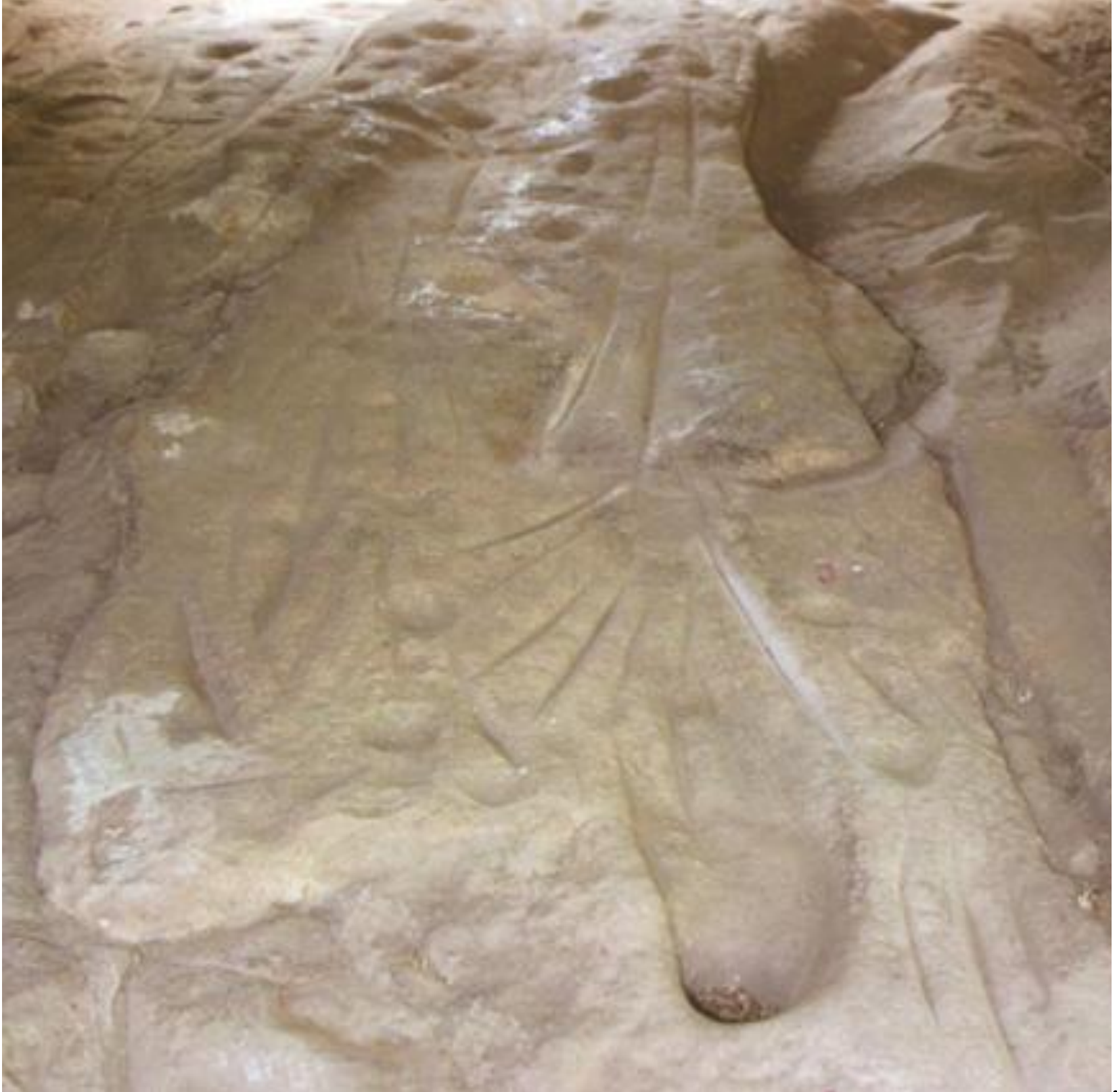

**Figure 8.** The biggest cup-shaped pit [21, fig. 58].

Eyewitnesses of Sikhote-Alin meteorite fall reported that in the bottom of the trajectory of fireball was already out of several parts and each part was moving on their own, leaving a narrow strip - a trickle of smoke trail. These trickles, expanding and merging together, formed one continuous dust trail [24, p. 60]. Therefore petroglyph claws paws we offer interpreted as smoky trickles of individual meteoroid fragments.

Several further into grotto depicts diverging fan-like lines, on the ends of which are stamped with small pits. Large cup-shaped pit with his lines - traces of an element of the fan too. Eyewitnesses of Sikhote-Alin meteorite fall seen that in the bottom of the meteoroid trajectory its main body - the fireball - accompanied by some small glowing satellites. Many witnesses reported that while driving meteoroid from it sparks flew. Several witnesses saw the meteoroid parts fall near the surface of the Earth [24, p. 60]. An eyewitness, who was at 9 km from place of fall, noticed crushing of meteoroid in the final part of its visible trajectory. He said that white-hot shards flew down tight group and fan-shaped [24, p. 62]. We think that the entire fan-shaped complex of petroglyphs depicts just such a crushing with meteoroid fragments flying apart like a fan in the final part of its trajectory.

Two almost parallel lines are drawn from the fan-shaped complex of petroglyphs in the depths of grotto in the side of the south entrance. They depict a trace of the meteoroid likely. A double line may indicate a double trail of smoke, such as at the Chelyabinsk meteoroid.

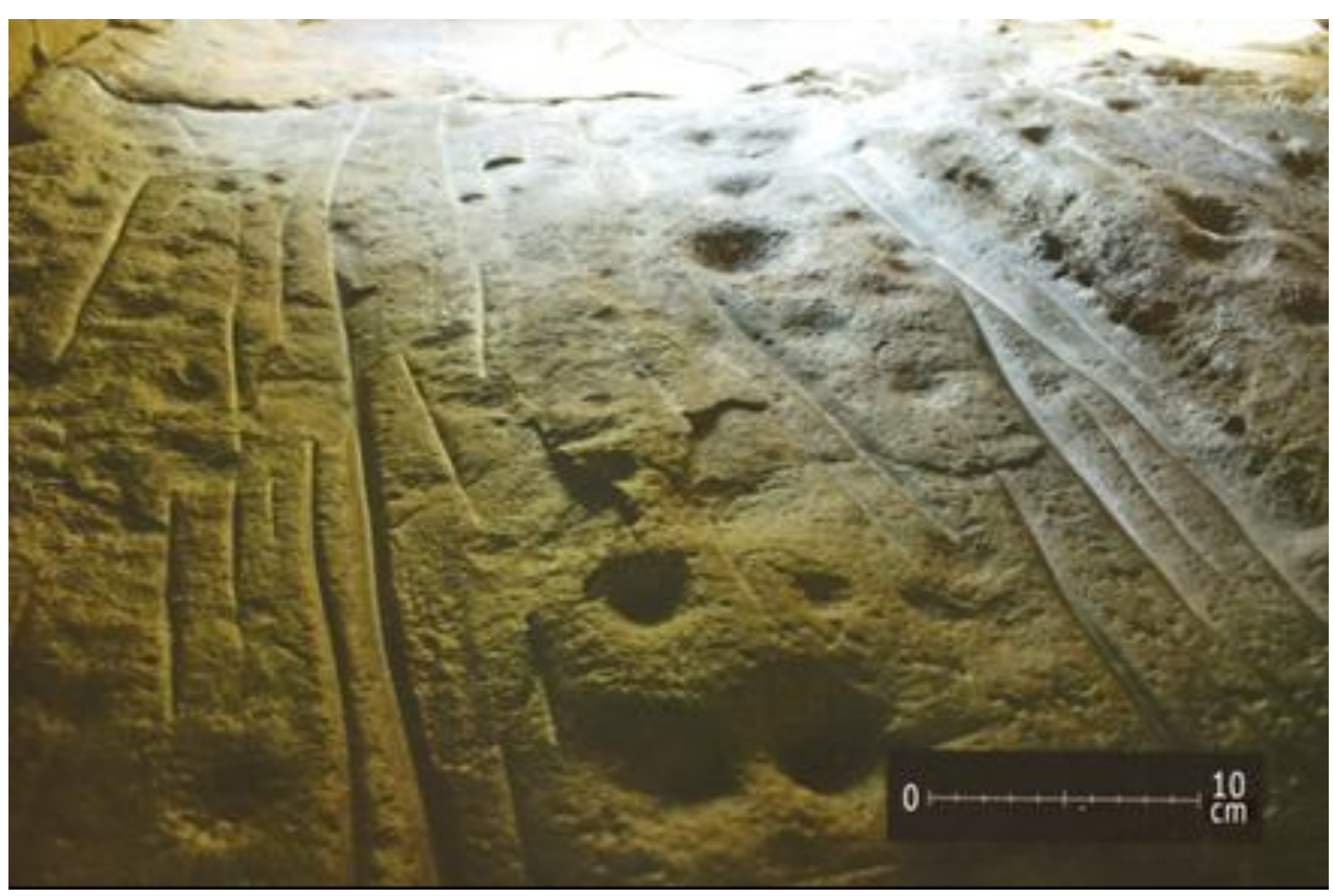

**Figure 9.** Petroglyphs in the central area of the quartz plate [21, fig. 16]

Lines and pits filled with all the central space of the quartz plate. The lines depicted, in the direction of the southern entrance to the northern entrance, approximately. The rows of pits are arranged in the same direction. From our point of view, these petroglyphs depict picture of the meteorite shower, in general. The lines represent the traces of meteoroid fragments in the sky, and pits - impact craters or shambles that followed the fall of the meteoroid and its fragments (Fig. 9).

Square petroglyphs we believe images of houses (Fig. 10, 11), and the rectangular petroglyphs are images outbuildings, such as corrals for livestock (Fig. 12).

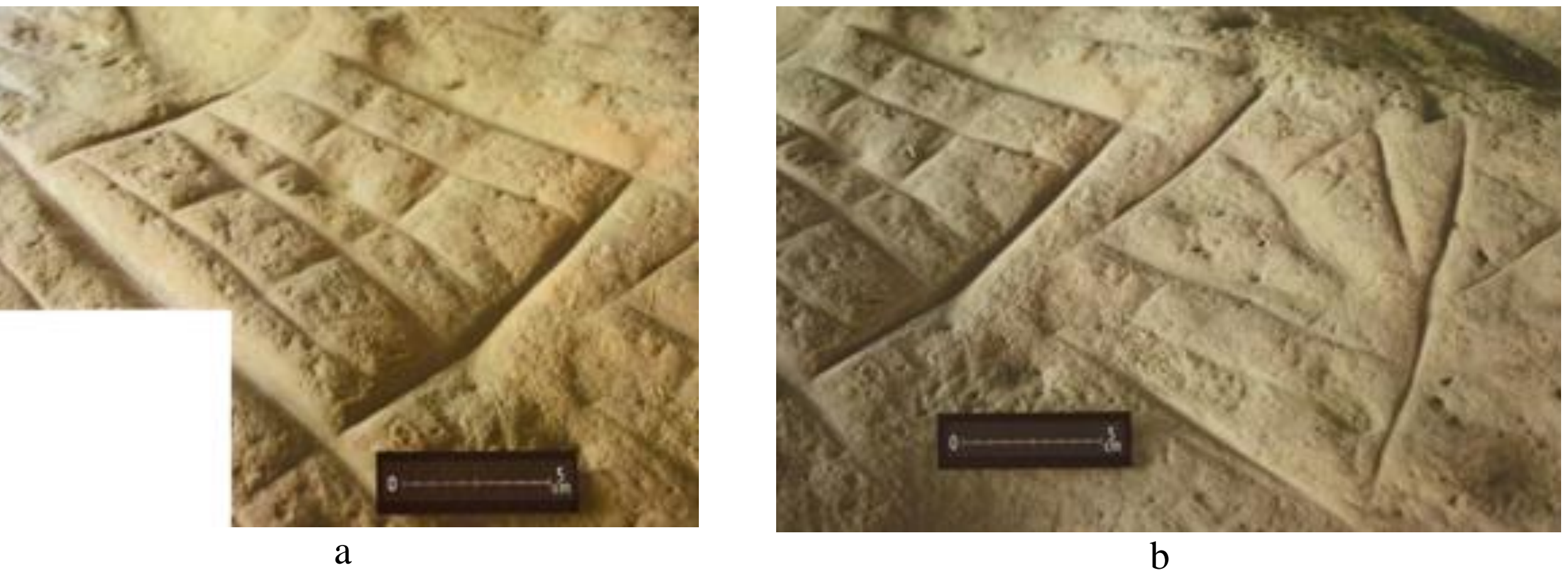

**Figure 10.** Large squares in the eastern area of the composition [21, fig. 48, 50].

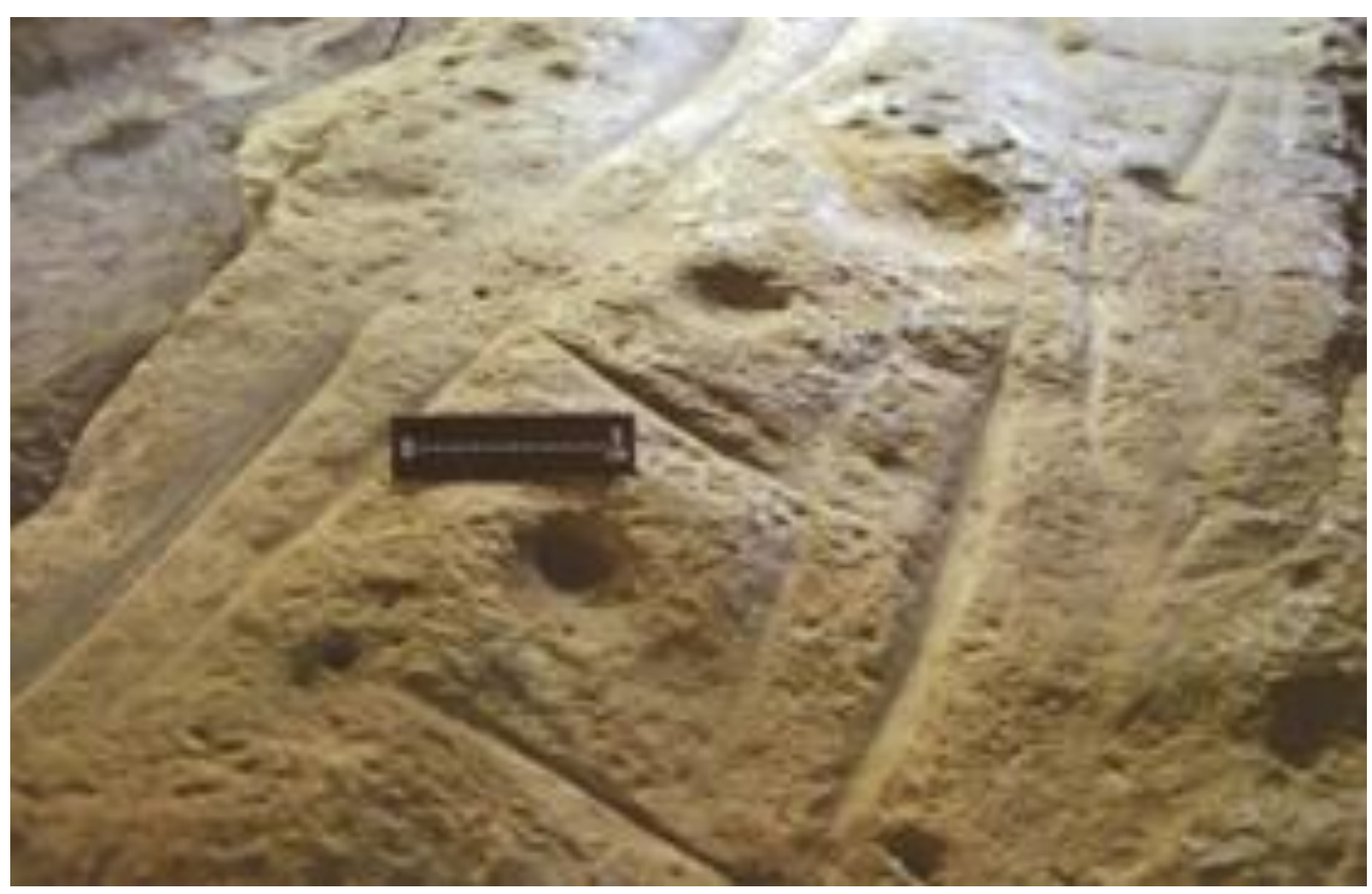

**Figure 11.** Small squares in the central area of the composition [21, fig. 54].

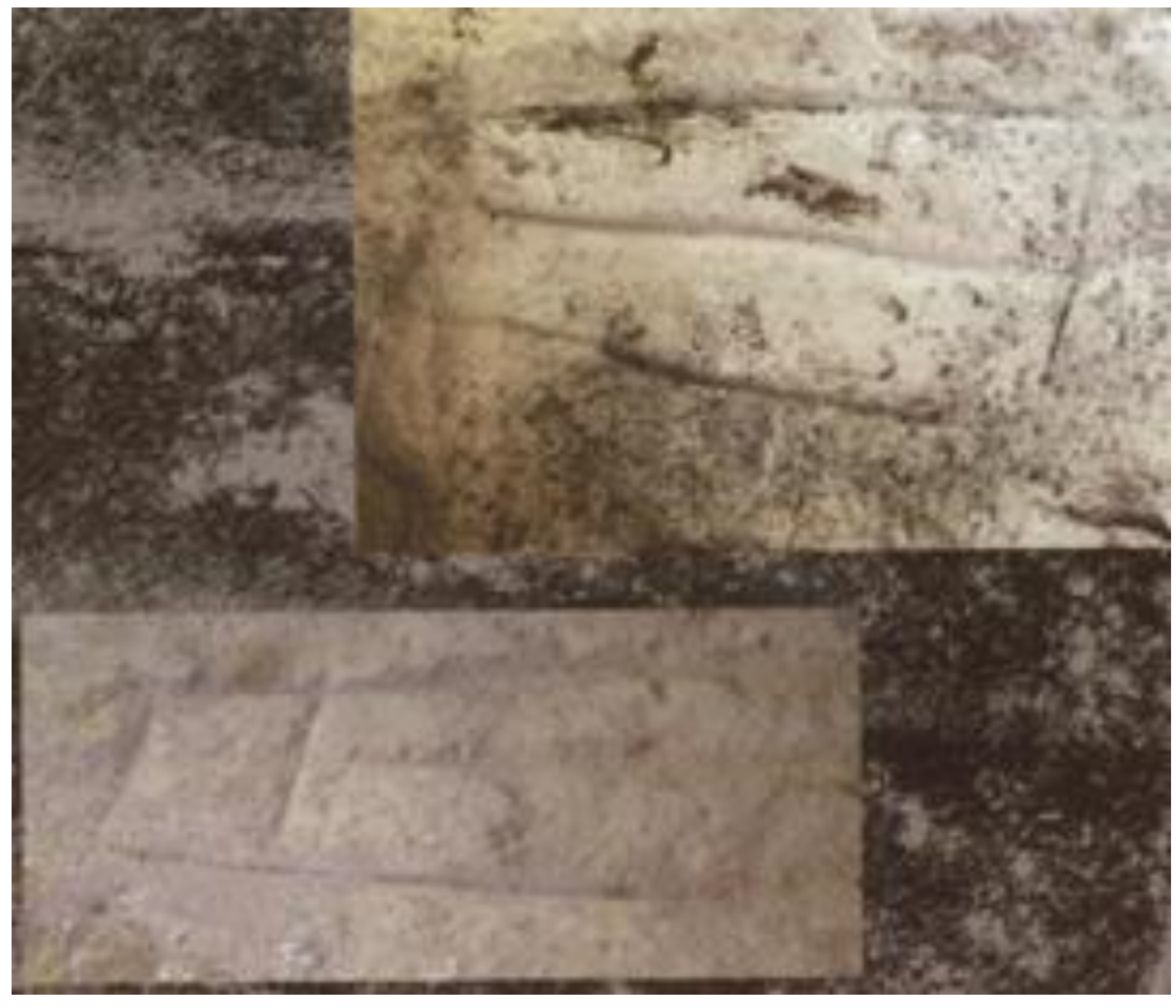

**Figure 12.** Rectangular petroglyphs [21, fig. 52].

The Yamnaya culture, to which Skelnovsky petroglyphs belong, was predominantly nomadic, with some agriculture practiced near rivers and a few hillforts. It occupied the territory from the Dniester to the Southern Urals and bordered on the cultures of North and Central Asia. Living and household buildings depicted with square and rectangular petroglyphs from the nomadic peoples of Asia in the Bronze Age, too (Fig. 13) [27].

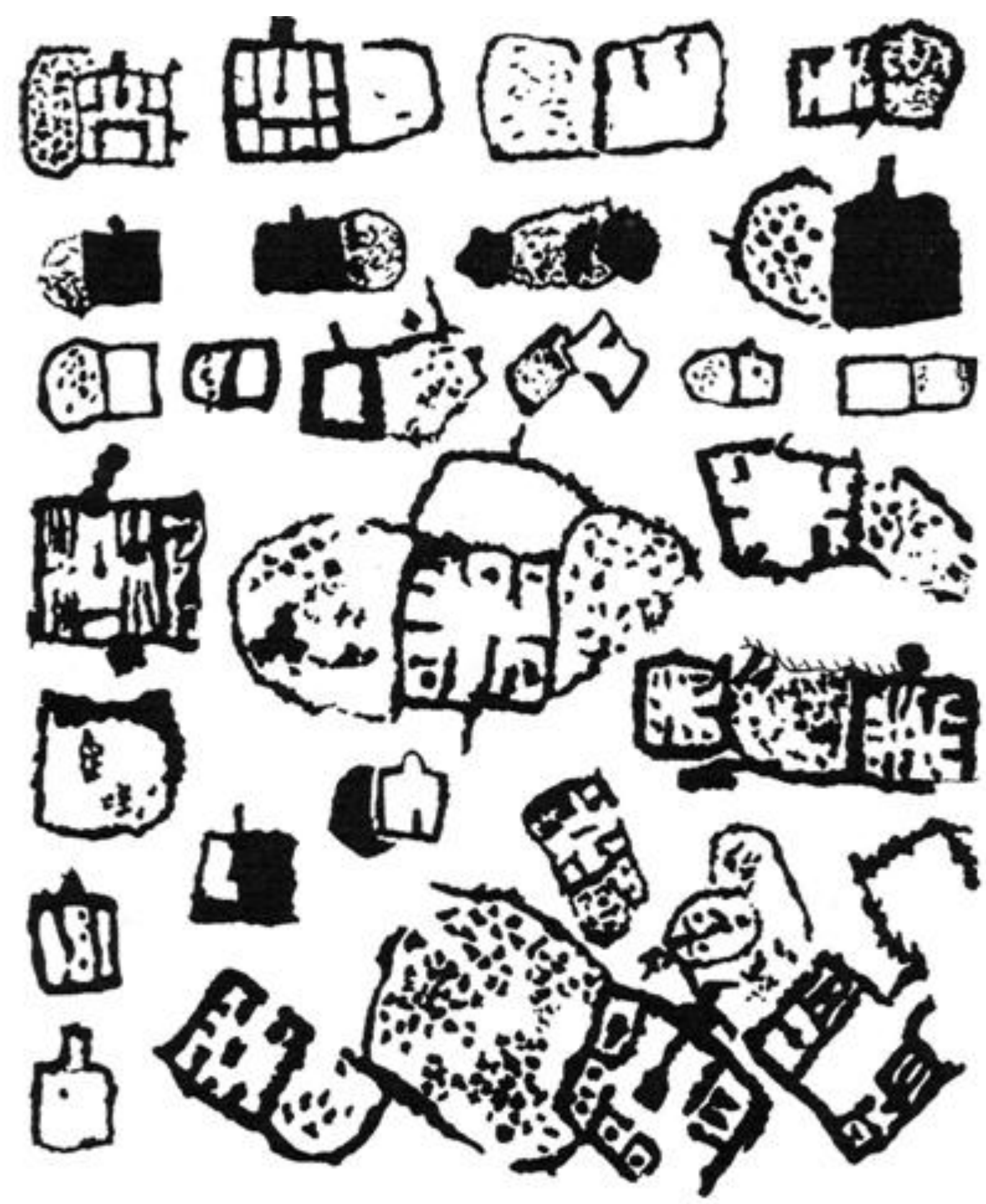

**Figure 13.** Mugur – Sargol, the Sayan Canyon of the Yenisei River. Images of houses and corrals for livestock petroglyphs [27, fig. 1].

We think that petroglyphs of the big squares may represent large wireframe houses of permanent settlements, and petroglyphs of small squares are temporary houses. The pits in the center of the squares, given the overall plot of composition, to symbolize the destruction of houses that have occurred from falling meteoroid fragments or blast wave is likely. Another petroglyph as a branch is located near the petroglyphs in the form of a large square image is partly in the eastern area of the composition. It can be interpreted as the image of second meteoroid crushing with flying apart of fragments that caused destruction of settlement houses (Fig. 10 b). In Skelnovsky grotto there is another type of petroglyphs - squares with marked diagonals. Small square with diagonals has a side length of about 4.5 cm (Fig. 14). The large square with diagonals, but them, has a lines parallel to its diagonals and divide them in half (Fig. 15). The large square is composed of small squares (with sides 4.5 cm) and their triangular halves so that ¼ of it diagonals is equal to ≈4.5 cm. This division of the large square allows, for example, it is easy to calculate its area, based on the area of the small square.

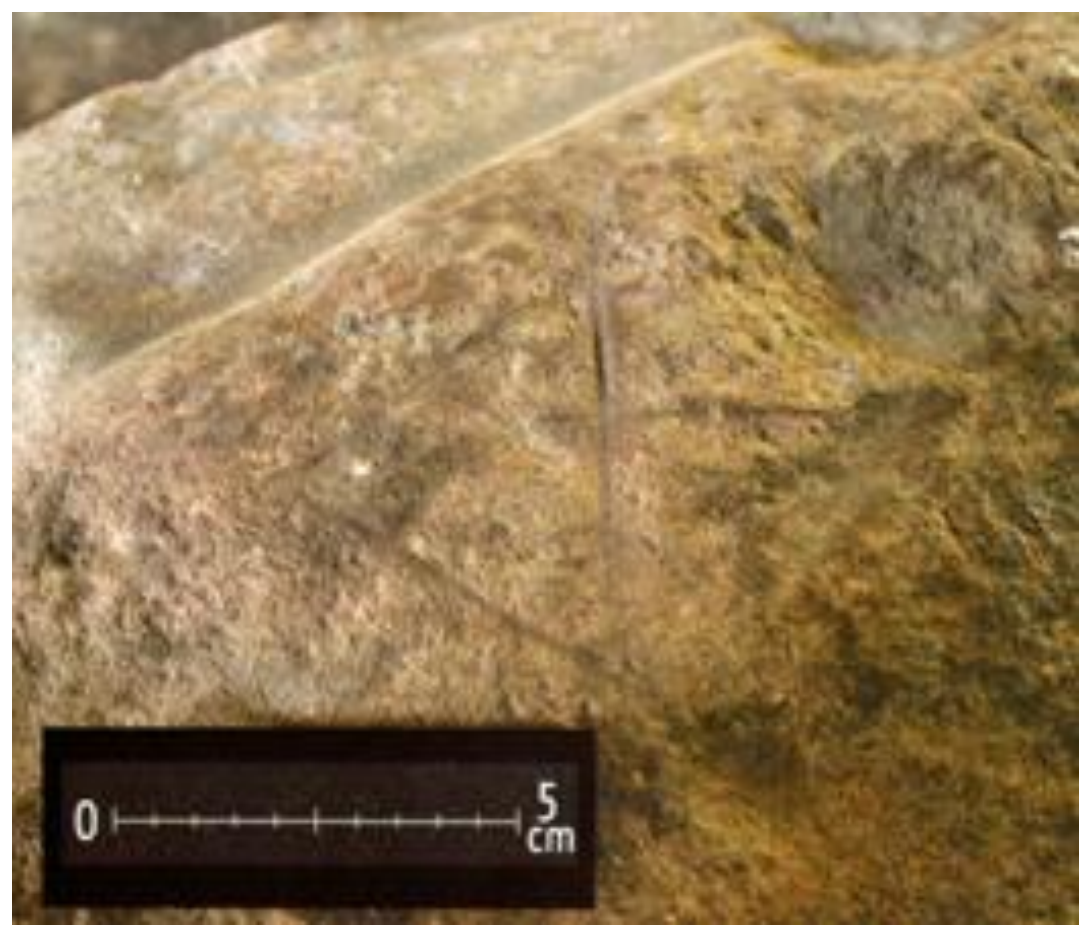

**Figure 14.** Petroglyph of small square with diagonals [21, fig. 28].

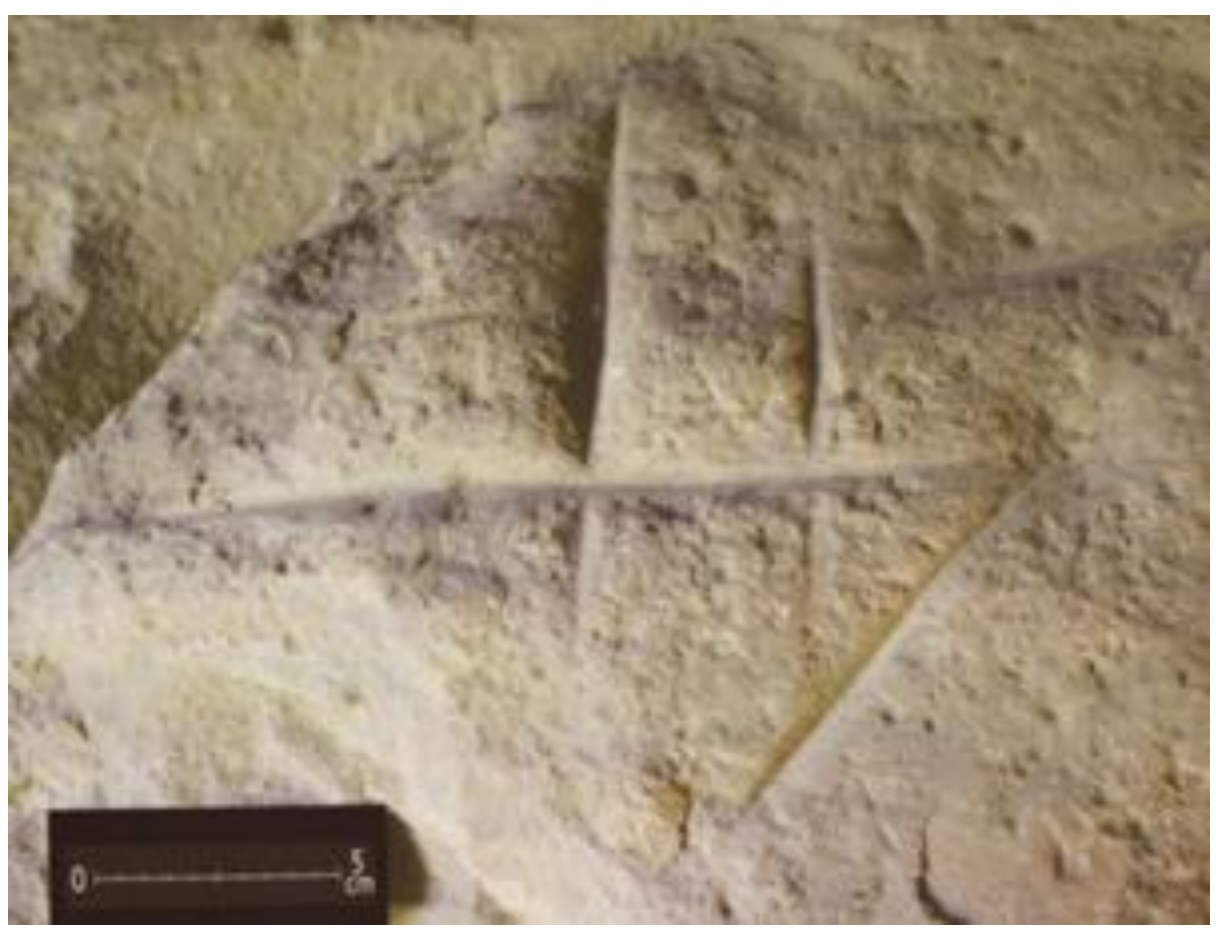

**Figure 15.** Petroglyph of the big square with diagonals [21, fig. 62].

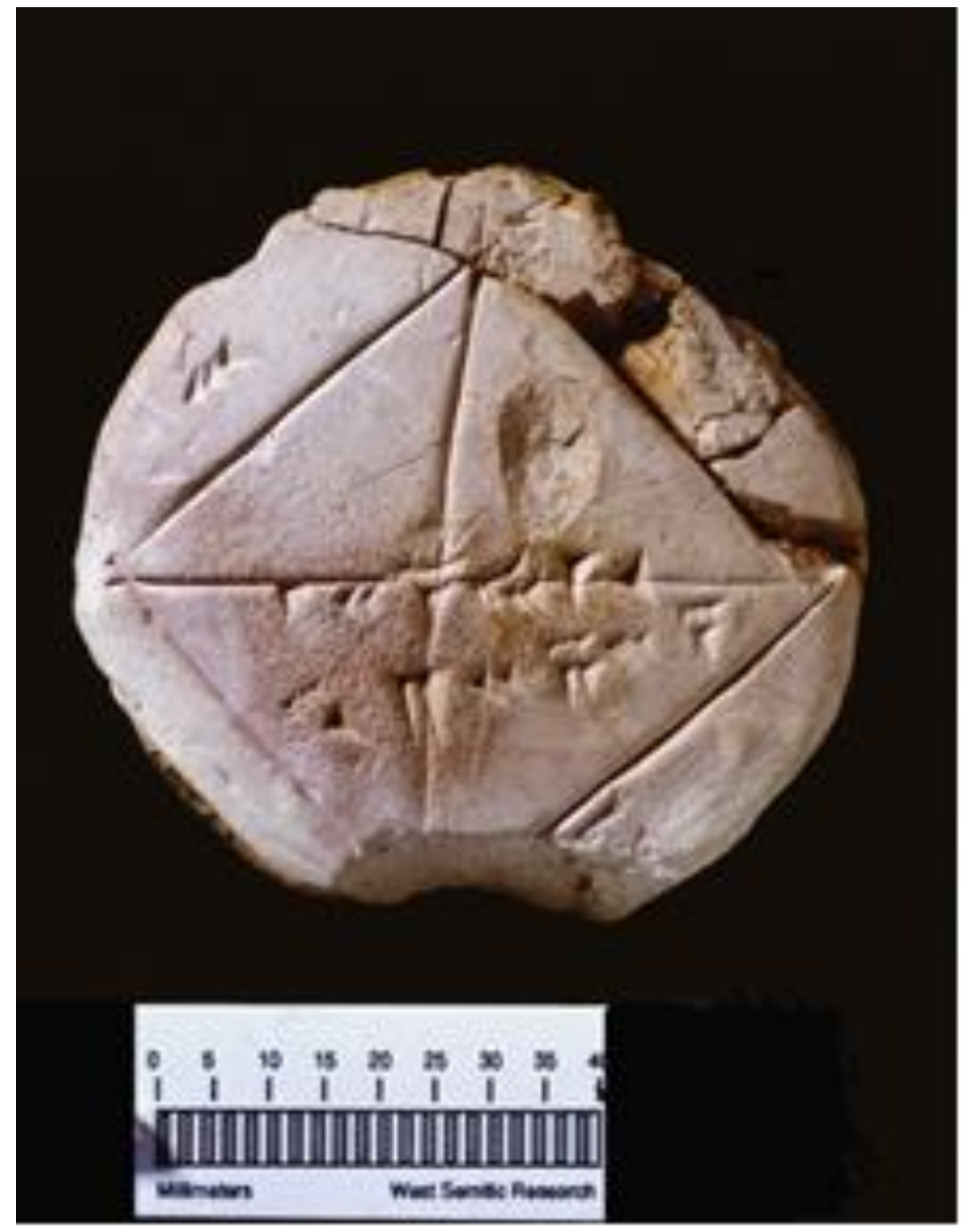

**Figure 16.** Old Babylonian clay tablet YBC 7289[2].

A very interesting fact is a coincidence Skelnovsky petroglyph small square sizes with the sizes of the square on the famous Old Babylonian clay tablet YBC 7289, dated 1800 - 1600 BC of the Yale Babylonian Collection (Yale Babylonian Collection (YBC)) [28] (Fig. 16). We believe that the Skelnovsky petroglyphs of squares with diagonals and the square on a clay tablet YBC 7289 could denote similar pyramidal constructions of an important economic or ritual purpose. The mathematical problem associated with such constructions, was captured on one of the Babylonian mathematical tablets that's why. Detection of images of similar constructions in the Northern Black Sea coast in the early Bronze Age could be due to the influence of

[2] http://mathdl.maa.org/mathDL/46/?pa=content&sa=viewDocument&nodeId=3889&pf=1

Mesopotamian cultures on Yamna culture mediated by the Maykop culture. Formation of the Maykop culture was directly related to the migration of individual groups of the north Mesopotamian population from the vast area stretching from the Tigris in the east to northern Syria and adjacent parts of eastern Anatolia in the west [29, p. 170]. Migrant population brought with him to the territory of the North Caucasus and Northern Black Sea is not only material things, but their technology, transferring impulse of cultural achievements in South-Eastern Europe from Western Asia. For example, the idea of pottery wheel has penetrated into the territory of the North Caucasus from Western Asia, where the circular pottery entered into the life of the urban civilizations in the III century BC already [29, p. 219]. It is possible that the discovered petroglyphs of squares with diagonals in the Skelnovsky grotto show a similar penetration into the Northern Black Sea region and the foundations protoscience knowledge from Western Asia.

**Reconstruction of the possible impact crater location**

Skelnovsky petroglyphs quite realistically portray the picture of meteoroid fall with the meteorite shower, from our point of view. The similarity of some of the elements of the composition of petroglyphs with eyewitness accounts of the Sikhote-Alin meteorite, suggesting not only a similar picture of the events, but also the similar nature of meteoroids. It can be attributed to the class of iron meteorites, just like the Sikhote-Alin meteorite, probably. Impact craters not been discovered in the Verkhnedonskoy district of Rostov region so far, however. The soils of the Verkhnedonskoy district been subject to intense wind and water erosion [30], and land of district are plowed for a long time, so small impact craters up to now have not been preserved most likely. Large craters, even if preserved, difficult to see in the modern landscape and still not identified as impact craters. We assumed that the petroglyphs were engraved on the plate in the grotto as realistically as possible and we compared the location of the largest cup-pit from a topographical map of the area[3]. Petroglyph of stone ax is oriented north-south approximately. It symbolizes the vertical falling presumably, so he was projected onto Skelnovsky grotto on the map. We took into account that the petroglyphs were engraved witness of meteoroid crushing in the final part of its trajectory within a radius of about 10 km, similar to the Sikhote-Alin meteorite. Therefore, we studied of topographic features on the satellite image and topographical map at a distance not exceeding 15 km from Skelnovsky grotto in the North-West direction (Fig. 17).

The magnetic declination $D$=8.17$^0$ $E$ for Skelnovsky grotto geographic coordinates $Lat$=49$^0$28$^/$ N and $Long$=40$^0$59$^/$ E for 2009. Magnetic declination was calculated using the program Magnetic declination online calculators (MDOC)[4] with an accuracy of 30$^/$. The program calculates magnetic declination using the model of International Geomagnetic Reference Field (IGRF), intended for the empirical representation of Earth magnetic field. Azimuth of meteoroid petroglyph is A≈-36$^0$ relative of stone ax petroglyph. Elliptical relief lowering found us through Google Earth 7.0.2 in this direction, on a distance of 14 km from the Skelnovsky grotto, approximately, next to Verhnyakovsky small village (Fig. 18 a).

---

[3] Topographical map M-37-94, 1:100000, ed. 1990, http://sunsite.berkeley.edu:8085/x-ussr/100k/M-37-094.jpg

[4] http://www.ngdc.noaa.gov/geomag-web/#declination

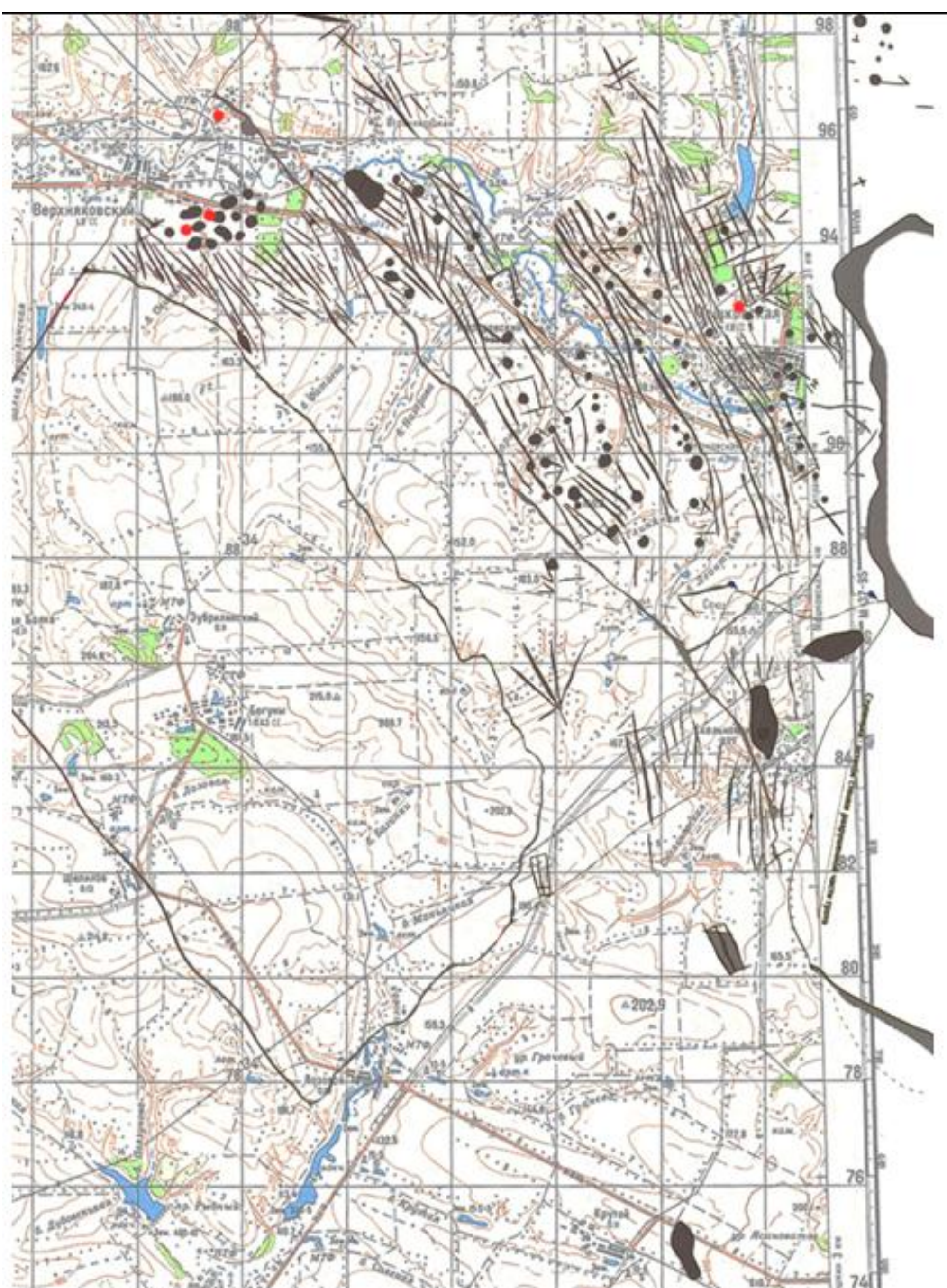

**Figure 17.** Projection of Skelnovsky petroglyphs on the topographical map. Red marked pits in the area of petroglyphs, marked on a topographical map.

We assume that this is a meteorite crater. Semi-minor axis of the ellipse has a length of ≈ 50 m, and the semi-major axis ≈ 160 m. The elliptical shape is typical for tangent impact craters. Rare semi-circular shape ravine is marked on a topographical map, located across the slope of Badorzhnaya beam (Fig. 18 b). It is located at a distance of about two kilometers from the alleged crater in the north-west of it, in the direction coinciding with the direction of meteoroid motion. On satellite images (Google Earth 7.0.2) can be clearly seen that the upper bound of the beam slope, coming from the semi-circular ravine in south-east direction, is very different from the surrounding terrain, due to the caving of the slope (Fig. 18 a). Caving of the beam slope can be explained by impact of the shock wave in the direction of meteoroid motion, produced them in the fall, from our point of view. We hope that this article will attract the attention of specialists, who will be able to organize an expedition for find and research of the meteorite crater and meteorite fragments in the Verhnyakovsky small village area.

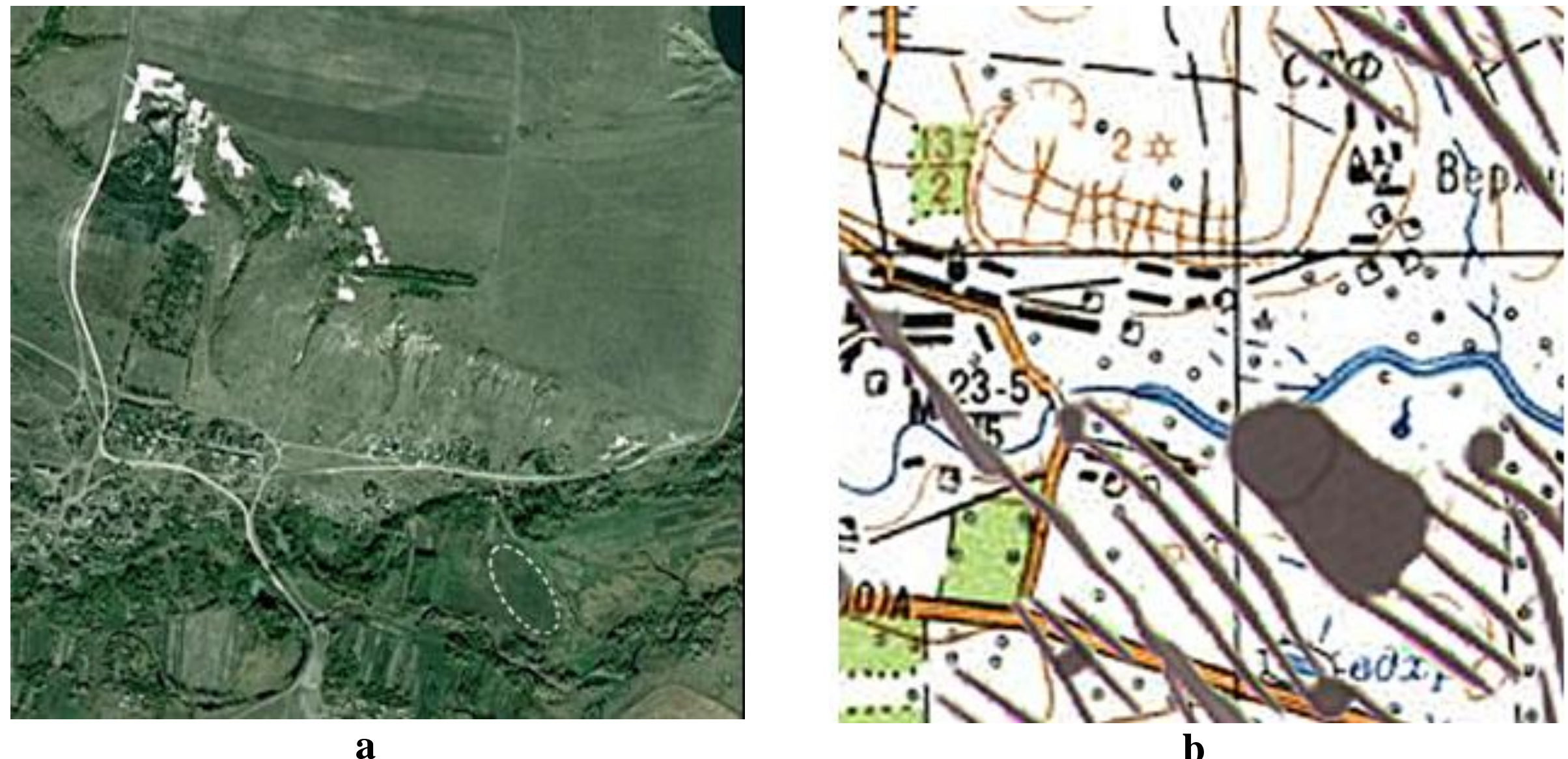

**Figure 18.** Expected meteoroid impact site: **a** - satellite photo (contour elliptic reduction terrain marked by a dashed line), **b** - fragment projection of Skelnovsky petroglyphs on the topographical map in area meteoroid petroglyph (semi-circular ravine in the upper left).

## Conclusions

The conclusion about astronomical nature of main content Skelnovsky grotto petroglyphs was made by us in this research. This conclusion was based on comparative analysis of contemporary eyewitness accounts of the Sikhote-Alin meteorite fall and Skelnovsky grotto petroglyphs, on the basis of modern astronomical knowledge about meteoroids nature and impact craters, on the basis of analysis of satellite images and topographical map of the Skelnovsky grotto region. Many petroglyphs can be associated with concrete phenomena, specific to the picture of meteoroid fall similar to the Sikhote-Alin meteorite fall, which was accompanied by a meteorite shower. Since the meteoroid falling and meteorite shower are colorful and frightening spectacle, especially when observers are not very far from place of fall, it is only natural that the ancient people has captured it in a rock petroglyphs.

Craters are not very young age and relatively small size, as in the case Skelnovsky meteoroid, found rarely. Therefore detection of the meteorite crater in the alleged place of Skelnovsky meteoroid falling would be of great importance not only for archaeoastronomy, but for meteoritics how section of modern astronomy.

## Acknowledgements

Authors of the article express their sincere gratitude to Doctor of Physics & Mathematics Y. A. Schekinov and Doctor of History V.Y. Kiyashko for the support of research.